\documentclass[superscriptaddress,aps,prb,reprint]{revtex4-1}
\usepackage{graphicx}
\usepackage{epstopdf}
\usepackage{braket}
\usepackage{mathrsfs}
\newcommand{\crmn}{Cr$_7$Mn}
\newcommand{\gazn}{Ga$_7$Zn}

\begin{document}

\title{A Clock Transition in the Cr$_7$Mn Molecular Nanomagnet}

\author{Charles A. Collett}
\author{Kai-Isaak Ellers}
\affiliation{Department of Physics and Astronomy, Amherst College, Amherst 01002, USA}
\author{Nicholas Russo}
\author{Kevin R. Kittilstved}
\affiliation{Department of Chemistry, University of Massachusetts, Amherst, MA 01003, USA}
\author{Grigore A. Timco}
\author{Richard E.P. Winpenny}
\affiliation{School of Chemistry, The University of Manchester, Manchester, UK}
\author{Jonathan R. Friedman}
\affiliation{Department of Physics and Astronomy, Amherst College, Amherst 01002, USA}
\date{\today}

\begin{abstract}
A viable qubit must have a long coherence time $T_2$. In molecular nanomagnets $T_2$ is often limited at low temperatures by the presence of dipole and hyperfine interactions, which are often mitigated through sample dilution, chemical engineering and isotope substitution in synthesis. Atomic-clock transitions offer another route to reducing decoherence from environmental fields by reducing the effective susceptibility of the working transition to field fluctuations.  The {\crmn} molecular nanomagnet, a heterometallic ring, features a clock transition at zero field. Both continuous-wave and spin-echo electron-spin resonance experiments on {\crmn}  samples diluted via co-crystallization, show evidence of the effects of the clock transition with a maximum $T_2\sim350$ ns at 1.8 K.  We discuss improvements to the experiment that may increase $T_2$ further.
\end{abstract}

\maketitle
The fundamental building block of a quantum computer is a qubit, a two-level system that can be in a superposition of the two levels. There are a wide variety of physical systems that can be used as qubits. Molecular nanomagnets (MNMs) have received significant attention due to their possible utility as spin qubits.\cite{leuenberger_quantum_2001,tejada_magnetic_2001,friedman_single-molecule_2010} MNMs can be chemically engineered, allowing the design of samples with desired characteristics, and manipulation and control of the spin state can be performed through electron-spin resonance (ESR).\cite{ardavan_engineering_2015,wedge_chemical_2012,timco_engineering_2009} The ideal qubit would feature a quantum state with a long lifetime that is able to be precisely controlled and measured, as well as coupled to an array of other such qubits in a controllable manner. The engineerability of MNMs affords the ability to couple multiple MNMs together to create multi-qubit systems.\cite{ardavan_engineering_2015,chiesa_molecular_2015,timco_linking_2011,ferrando-soria_modular_2016} In order for such systems to be viable qubits, the coherence time $T_2$ needs to be long compared to the time required for quantum gates. Here we present experimental results showing an enhancement of $T_2$  in the {\crmn} MNM through the use of a so-called atomic-clock transition.

\begin{figure}[b!]
\includegraphics[width=2.5in]{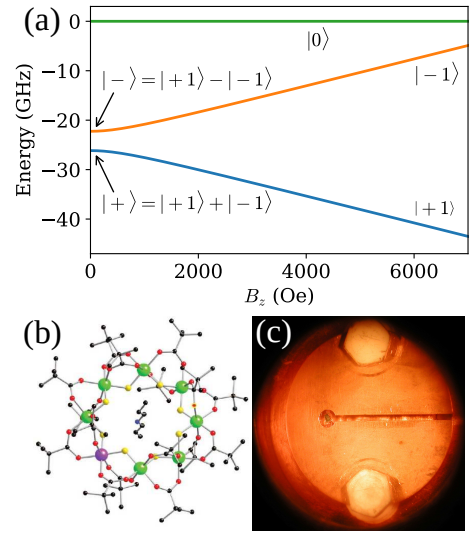}
\caption{\label{fig:monomer}(a) Energy-level diagram for a {\crmn} molecule, showing the zero-field avoided crossing between $\ket{m=\pm1}$ states, creating the $\ket{\pm}$ clock states. (b) Molecular structure of {\crmn}. (c) Picture of a loop-gap resonator with a {\crmn} sample in the loop.}
\end{figure}

{\crmn} is a paramagnetic ring [(CH$_3$)$_2$NH$_2$][Cr$_7$MnF$_8$((CH$_3$)$_3$CCOO)$_{16}$] (``\crmn''), shown in Fig.~\ref{fig:monomer}(b).\cite{larsen_synthesis_2003} At low-temperature it has a ground-state spin of $S=1$ that can be described by the spin Hamiltonian
\begin{equation}
  \mathscr{H}=-DS_{z}^2 + E(S_{x}^2-S_{y}^2)-g\mu_B\vec{B}\cdot\vec{S}, \label{eq:SHam}
\end{equation}
where $D=21$ GHz, $E=1.9$ GHz, $g=1.96$ 
is the Land\'e $g$-factor and $\mu_B$ is the Bohr magneton. The $D$ term describes the longitudinal anisotropy of the molecule, and gives the $|\!\pm\!1\rangle$ states a lower energy than the $|0\rangle$ state, while the $E$ term describes the transverse anisotropy that breaks the zero-field degeneracy of the $|\!\pm\!1\rangle$ states by mixing them, as shown at the left edge of Fig.~\ref{fig:monomer}(a). Due to this mixing, the two lowest-energy levels are the superpositions $|\pm\rangle=(|\!+\!1\rangle\pm|\!-\!1\rangle)/\sqrt{2}$, forming an avoided crossing at zero field with a ``tunnel splitting'' energy $\Delta=2E$. The application of a magnetic field along the easy axis, $B_z$, results in an additional Zeeman splitting of the two lowest energy levels with no change in the energy of the $|0\rangle$ state, as shown in Fig.~\ref{fig:monomer}(a).

Near the zero-field avoided crossing, the leading-order dependence of the tunnel splitting on the magnetic field is quadratic: $\Delta\propto B_z^2$. Because of this dependence, at zero field the splitting is, to first order, insensitive to field fluctuations: $\frac{d\Delta}{dB_z}\big|_{B_z=0}=0$. This type of transition is known as an atomic-clock, or just clock, transition, based on its use in the atomic clock community to achieve frequency stability.\cite{bollinger_laser-cooled-atomic_1985} A recent experiment in a Ho-based MNM has shown that clock transitions produce significant increases in quantum coherence times,\cite{shiddiq_enhancing_2016} as the main source of decoherence at low temperatures is dipolar interactions with neighboring spins, whose fluctuations cause changes in the local magnetic field seen by the central spin.\cite{takahashi_decoherence_2011} At a clock transition, the effect of these fluctuations is suppressed, leading to the observed increase in coherence times. Here we show evidence of the effect of a clock transition in {\crmn}.


We use electron spin resonance to probe and control the state of this molecule. As the effect of the clock transition is strongest right at the zero-field tunnel splitting, we need to precisely tune our radiation frequency to the splitting energy. We do this by using custom loop-gap resonators (LGRs), which can be easily designed and then tuned to precise frequencies by inserting a dielectric into the gap. LGRs concentrate the oscillating magnetic field in a narrow region, producing high filling factors.\cite{froncisz_loop-gap_1982} Our sample sits in the loop of the resonator, as shown in Fig.~\ref{fig:monomer}(c), and we couple radiation into the system through antennae above the loop and/or the gap. Continuous wave (CW) measurements were done using a vector network analyzer (VNA, KeySight E5063A), and pulsed measurements were done using a homemade spectrometer.

\begin{figure}[t!]
	\includegraphics[width=2.5in]{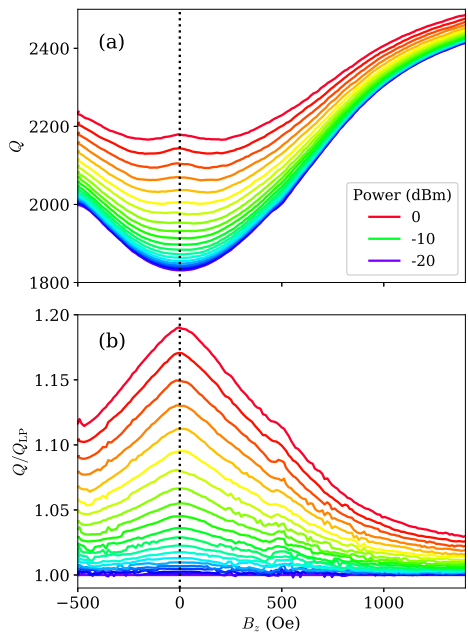}
	\caption{\label{fig:cw_decoupling}(a) Resonator $Q$ vs.~applied magnetic field at 4.00~GHz and powers from -20 to 0~dBm. (b) Resonator $Q$ divided by the $Q$ at -20~dBm to emphasize the decoupling effect. The sharp peak around zero field suggests this decoupling is associated with the clock transition.}
\end{figure}
For our CW measurements, we put a crystal of {\crmn} in the loop of an LGR tuned to 4.00~GHz, cooled it to 1.8 K, and then swept a magnetic field applied along the easy axis while measuring $Q=\nu/\Delta\nu$ of the resonator, where $\nu$ is the resonance frequency and $\Delta\nu$ is the full width at half maximum of the resonance peak. When the sample is on resonance with the resonator, it absorbs energy from the rf magnetic field in the loop, increasing the loading and reducing the $Q$ of the resonator. As such, we measure the coupling of the sample to the resonator by  reduction in  $Q$.

Fig.~\ref{fig:cw_decoupling}(a) shows measured values of $Q$ as a function of magnetic field. The lowest (purple) trace corresponds to an input power of -20~dBm, while our highest power of 0~dBm is red, with intermediate powers designated with intermediate spectral colors. The slight dip around 500~Oe is due to an impurity unrelated to the sample. At the lowest powers we see a broad dip in $Q$ around zero field, as expected for a transition between the two levels in the avoided crossing (cf.~Fig.~\ref{fig:monomer}(a)); the width of the peak is due to significant inhomogeneous broadening in the sample, as well as the fact that the transition frequency depends quadratically on field near the clock transition. As we increased the power,  $Q$ also increased over the entire field range, but with a pronounced shape change near zero field. An increase in $Q$ indicates a decrease in the coupling between sample and resonator, so this effect shows that as power increases the sample is decoupling from the resonator preferentially around zero field.

To isolate this power dependence, we divided the -20~dBm $Q$ data from the higher power $Q$ data,  shown in Fig.~\ref{fig:cw_decoupling}(b) as $Q/Q_{\rm{LP}}$. The resulting ``decoupling peak'' is nearly triangular, being sharply peaked at zero field; the shape change seen in the raw $Q$ data is the result of this triangular decoupling peak superimposed with the smoother resonance dip. This plot shows that the decoupling has a smooth, monotonic power dependence and is strongest near zero field. The reduction in coupling with increasing power is consistent with saturation of the transition, but the shape of the decoupling peak suggests that the saturation is being strongly enhanced near zero field. As the main feature at zero field is the avoided crossing, we propose that this decoupling peak demonstrates saturation enhanced by the increase in $T_2$ produced by the clock transition.

While our CW measurements suggest the existence of a clock transition in {\crmn}, to verify it we directly measured $T_2$ using a homemade pulsed-ESR spectrometer, configured to perform a Hahn echo sequence. Two RF pulses are sent to the resonator, the first with a duration such that it will tip the spin state by an angle of $\pi/2$, followed after a time $\tau$ by a second pulse which tips the state by an angle of $\pi$, producing a ``spin echo'' a time $\tau$ after the second pulse, resulting in the sequence $\pi/2-\tau-\pi-\tau-$echo. Fitting the dependence of the echo signal on $2\tau$ to an exponential decay gives a time constant of $T_2$. In practice, we also performed background subtraction in one of two ways. For our preliminary experiments we took data at a low field and a much higher field, where the sample is no longer coupled to the resonator; subtracting the two signals removes the background. For our $T_2$ experiments we alternated sending in a $\pi$ pulse several $\mu$s before the $\pi/2$ pulse.  This extra pulse inverts the echo; subtracting signals with this extra pulse from signals without the extra pulse removes the background while effectively doubling our echo signal.

Using the same {\crmn} crystal from our CW experiments we were unable to see any echo, indicating that $T_2$ is too short for our apparatus to detect because of strong dipolar interactions between molecules within the crystal. To increase $T_2$, we diluted the {\crmn} molecules, increasing the average distance between molecular spins,\cite{henderson_control_2008,moro_coherent_2013} through co-crystallization  with an isostructural diamagnetic analogue, [(CH$_3$)$_2$NH$_2$][Ga$_7$ZnF$_8$((CH$_3$)$_3$CCOO)$_{16}$] (``\gazn''). The resulting solid solution crystal has the advantage of preserving the orientational order of the molecules. Previous measurements of dilute {\crmn} suggest a maximum coherence time of $\sim600$~ns at 2~K in the dilute limit, increasing to $\sim4~\mu$s for deuterated samples.\cite{ardavan_will_2007}

\begin{figure}[b!]
	\includegraphics[width=2.5in]{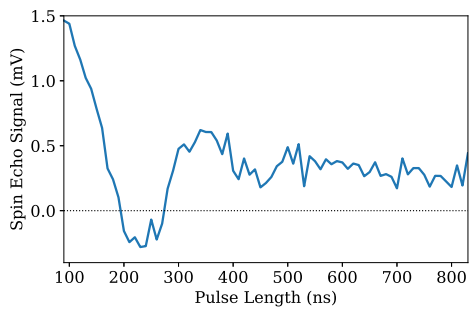}
	\caption{\label{fig:nutation}Spin echo signal as a function of the length of the first pulse. The resulting Rabi oscillations confirm that our signal comes from spin echo. The offset is an artifact of our background-subtraction method.}
\end{figure}
With samples volumetrically diluted to 10\% and 0.5\% {\crmn} in {\gazn}, $T_2$ was long enough to be measurable. To verify that we were observing a spin-echo signal, we performed nutation measurements by changing the length of the extra pulse. Some results for a 10\% sample, background-subtracted using the high field method described above, are plotted in Fig.~\ref{fig:nutation}, showing the expected Rabi oscillations. We used these results to determine the starting duration of our $\pi/2$ and $\pi$ pulses, which we then tuned to maximize our signal.

\begin{figure}[b!]
	\includegraphics[width=2.5in]{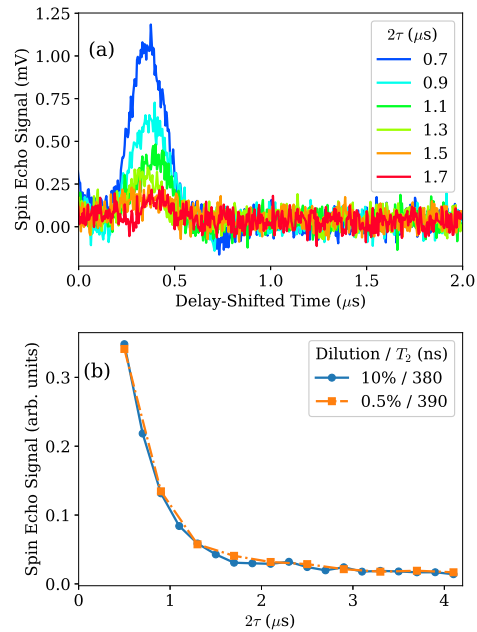}
	\caption{\label{fig:se_gazn}(a) Background-subtracted spin echo signal at various delay times $2\tau$. The time axis has been shifted such that every echo occurs at the same effective time. (b) Spin echo peak height as a function of the delay time $2\tau$ for 10\% (blue circles) and 0.5\% (orange squares) samples. The lines connecting datapoints are guides to the eye.}
\end{figure}
Spin echo results for a 10\%-dilution sample at several different delay times $\tau$, background subtracted using the third-pulse method described above, are shown in Fig.~\ref{fig:se_gazn}(a), with each trace offset in time so the echoes occur at the same place in the plot. We measured $T_2$ by plotting the echo signal area
vs.~$2\tau$, as shown in Fig.~\ref{fig:se_gazn}(b), and fitting to an exponential decay function to obtain $T_2=340$~ns at an applied field of 20 Oe
for the 10\% sample. Surprisingly, our measured $T_2$ value for the 0.5\% sample was only slightly higher, $T_2=360$~ns at 20 Oe.
Thus, different dilution levels did not produce the expected increase in $T_2$, instead yielding roughly constant coherence times. As we are near neither the dilute limit nor the expected maximum $T_2$, we suspect that the co-crystallization method is limiting our value of $T_2$ somehow, either through spectral diffusion produced by spins in the {\gazn}, interaction with the nuclear spin of Ga, or acoustic phonons in the crystal. Alternatively, the {\crmn} may be aggregating in the co-crystallization, limiting the effect of dilution. The presence of an echo signal for the dilute samples shows that dilution is having some effect, but the lack of change in $T_2$ indicates that the effect of this method of dilution for this molecule has saturated by the time we have reached 10\% dilution. This limitation should be specific to our co-crystallization method, and we expect the previously-used method of diluting {\crmn} by dissolving it in toluene should be more effective, as that technique has yielded larger values of  $T_2$.\cite{ardavan_will_2007} Experiments to create such dilute solutions of {\crmn} and measure their properties are ongoing. Characterization of such samples will allow us to determine if the field dependence of $T_2$ exhibits the expected enhancement near zero field, the center of the avoided crossing.

\begin{acknowledgments}
We thank G. Joshi, S. Hill, A. Lupascu, J. Baugh and M. Blencowe for useful conversations, and C. Yoo and E. Turnbull for preliminary sample characterization work. Support for this work was provided by the U.~S.~National Science Foundation under Grant Nos.~DMR-1310135 and DMR-1708692 and by the Amherst College Dean of Faculty. The research was also funded by the EPSRC(UK) via an Established Career Fellowship (EP/R011079/1) to R.E.P.W.
\end{acknowledgments}

\bibliography{CWCT}{}

\end{document}